\documentclass[10pt,notitlepage,showpacs,pra,twocolumn]{revtex4-1}
\usepackage{ulem}
\usepackage{amsfonts}
\usepackage{amsmath}
\usepackage{amssymb}
\usepackage{graphicx}
\usepackage{float}
\usepackage{color}
\usepackage[toc,page,header]{appendix}

\begin{document}

\title{Unidirectional and controllable higher-order diffraction by a Rydberg electromagnetically induced grating}
\author{Dandan Ma$^{1}$, Dongmin Yu$^{1}$, Xingdong Zhao$^{2}$ and Jing Qian$^{\dagger,1}$ }
\affiliation{$^{1}$State Key Laboratory of Precision Spectroscopy, School of Physics and Material Science, East China
Normal University, Shanghai 200062, China}
\affiliation{$^{2}$College of Physics and Materials Science, Henan Normal University, Xinxiang 453007, China}

\begin{abstract}
A method for diffracting the weak probe beam into unidirectional and higher-order directions is proposed via a novel Rydberg electromagnetically induced grating, providing a new way for the implementations of quantum devices with cold Rydberg atoms. The proposed scheme utilizes a suitable and position-dependent adjustment to the two-photon detuning besides the modulation of the standing-wave coupling field, bringing a in-phase modulation which can change the parity of the dispersion. We observe that when the modulation amplitude is appropriate, a perfect unidirectional diffraction grating can be realized. In addition, due to the mutual effect between the van der Waals (vdWs) interaction and the atom-field interaction length that deeply improves the dispersion of the medium, the probe energy can be counter-intuitively transferred into higher-order diffractions as increasing the vdWs interaction, leading to the realization of a controllable higher-order diffraction grating via strong blockade.
\end{abstract}
\email{jqian1982@gmail.com}
\maketitle
\preprint{}

\section{Introduction}
In the field of quantum simulation, designing controllable quantum devices such as quantum gate, quantum annealer based on a cold atomic medium has acquired significant progress, mainly because of long coherence time and flexible manipulation possessed by an atom-field interacting system in a low temperature \cite{Bloch08}. 
An ultracold neutral atomic source can be used to realize a robust quantum simulator taking advantage of their internal hyperfine levels serving as qubits \cite{Briegel00,Weiss17}, providing further operations for multi-particle entanglement \cite{Mandel03} and fast quantum gate \cite{Jaksch00}.
Undoubtedly, highly-excited Rydberg atom as one of the neutral atoms, has manifested as an attractive candidate for persisting the coherence and for the realization of new quantum devices in quantum simulation field, applying for such as quantum simulator in a spin model by the strong many-body interactions \cite{Weimer10}, the controlled high-fidelity entanglement with a reduced-phase-noise laser \cite{Levine18} and so on.

In parallel, electromagnetically induced transparency (EIT) \cite{ Fleischhauer05} plays a significant role for studies of optical devices in an atomic medium, offering great advances for nonlinear quantum optics \cite{Firstenberg16}. EIT essentially utilizes quantum interference of double optical transitions to vanish the absorption of the weak probe field, resulting in a EIT window to enhance the probe transmission even in the case of resonant probe detuning. 
In a Rydberg system, by coupling the probe transition to a Rydberg state via EIT, the strong van der waals(vdWs) interactions between two Rydberg states can be translated into sizable interactions between photons, resulting in cooperative optical nonlinearity \cite{Pritchard10}. Besides, EIT has been applied to store the gate photon as a Rydberg excitation, realizing various quantum devices such as single-photon switches \cite{Baur14,Murray16} or transistors \cite{Gorniaczyk14,Gorniaczyk16}.

It is remarkable that, in a Rydberg-EIT system, when the strong coupling field is replaced by a standing-wave(SW) field, implementing a spatial periodic modulation for the absorption (amplitude) and dispersion (phase) of medium, the traveling-wave (TW) probe field can be diffracted into higher-order directions. This is named as a Rydberg electromagnetically induced grating (Rydberg-EIG), serving as a new member in the family of Rydberg quantum devices. A normal atomic EIG, first proposed by Ling \cite{Ling98} and observed by Mitsunaga in sodium atoms \cite{Mitsunaga99}, has been widely explored {\it e.g.} \cite{Cardoso02,Brown05,Dutta06,Carvalho11,Liu16}. Other schemes created diffraction gratings based on the modulation of Raman gain without EIT tools \cite{Kuang11,Arkhipkin18}. Here differing from the normal ones, the proposed Rydberg-EIG with the uppermost level replaced by a Rydberg level is significantly influenced by the vdWs interaction. Intuitively, the diffraction intensity will exponentially decrease with the increase of vdWs interaction due to the breakup of EIT condition, representing no notable results \cite{Asghar16}.

Motivated by a recent work \cite{Liu17} where authors exploited an asymmetric EIG with parity-time symmetry to the coupling field breaking the parity of absorption, and realized to diffract the probe field into either negative or positive angles, we propose a new approach for achieving an exotic Rydberg-EIG with perfect unidirectional and higher-order diffraction. The key lies in introducing a suitable periodic modulation to the two-photon detuning in order to break the parity of dispersion, arising a position-dependent modulation to the energy of Rydberg level aside from the vdWs shift. As a result, we observe a perfect unidirectional diffraction with the peak $n$th-order diffraction intensities satisfying $I_p^{pk}(\theta_0)>I_p^{pk}(\theta_{-1})>I_p^{pk}(\theta_{-2})>I_p^{pk}(\theta_{-3})$, under the condition of an appropriate modulation amplitude of detuning. More interestingly, the diffraction intensity represents a strong oscillatory behavior and are found to be transferred into higher-order diffractions by increasing the strength of vdWs interaction, leading to a controllable enhanced higher-order diffraction grating via an optimal control for the vdWs interaction. Our scheme is of significant interest for scientists to design various quantum devices such as a large-angle optical splitter, with Rydberg atoms.

 .

\section{atom-field interaction model}

\begin{figure}
\centering
\includegraphics[width=3.4in,height=1.55in]{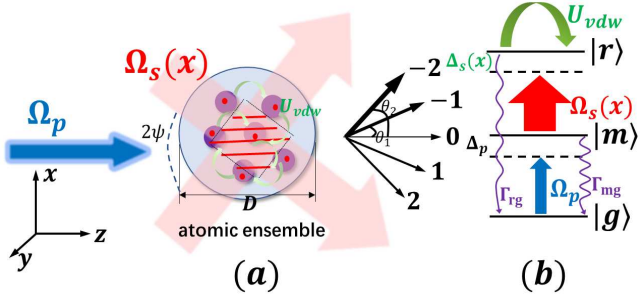}
\caption{(color online). (a) Schematic representation of a normal Rydberg-EIG. A single atomic ensemble with cascade three-level atoms is coupled by a strong SW coupling field and a weak TW probe field. When the incident probe field normally propagates through such an ensemble it can be diffracted into high-order diffractions due to the phase and amplitude modulations exerted by the SW coupling field. (b) Atomic energy levels. The vdWs-type interaction $U_{vdW}$ presents if atoms are simultaneously excited to the Rydberg state $\left\vert r\right\rangle$ due to the imperfection of full blockade. Relevant parameters are described in the text.
 } 
\label{model}
\end{figure}

We consider that a single ensemble with $N$ atoms has cascade three-level structure, see Fig. \ref{model}, driven by a weak TW probe field and a strong SW coupling field, whose Rabi frequencies are denoted as $\Omega_p$ and  $\Omega_s(x) = \Omega_{s0}\sin(\frac{\pi x}{\Lambda_{sx}})$. Here $\Omega_p$, $\Omega_{s0}$ are the peak amplitudes and $\Lambda_{sx} = \lambda_s/\sin\psi$ is the SW spatial period on $x$-axis with $\lambda_s$ the coupling wavelength and $\psi$ the azimuth. The energy level of each atom is composed of a Rydberg state $\left\vert r\right\rangle$, an intermediate excited state $\left\vert m\right\rangle$ and a ground state $\left\vert g\right\rangle$, respectively making $\left\vert g\right\rangle\to\left\vert m\right\rangle$ and $\left\vert m\right\rangle\to\left\vert r\right\rangle$ transitions. Note that, in a normal EIG, the strong SW field $\Omega_s(x)$ will exert a spatial periodic change to the refractive index of the atomic medium, leading to a far-field Fraunhofer diffraction to the weak probe field as it propagates through the medium \cite{Ling98}, and it becomes a Rydberg-EIG when the uppermost level is a Rydberg state \cite{Bozorgzadeh18}.

Actually, the typical timescales for Rydberg experiments is maximally on the order of Rydberg lifetime $\sim \mu$s during which the ultracold atoms moves only small distances relative to their separations, leading to a negiligence of the atomic motion in the frozen-gas environment \cite{Anderson98,Low12,Browaeys16}.
Here we assume that the interatomic interaction between two Rydberg atoms to be a vdWs-type \cite{Beguin13}, therefore the Hamiltonian in the frame of rotating-wave approximation can be written as $\mathcal{H} = \mathcal{H}_a+\mathcal{U}_{af}+\mathcal{U}_{vdW}$, which consists of an unperturbed atomic part $\mathcal{H}_a = -\hbar\sum_{j}^{N}[\Delta_p\sigma_{mm}^{j} + \Delta_s\sigma_{rr}^{j}] $, an atom-field coupling part $\mathcal{U}_{af} = -\hbar\sum_{j}^{N}[\Omega_p\sigma_{mg}^{j}+\Omega_s(x)\sigma_{rm}^{j}+H.c.]$ and an interaction part $\mathcal{U}_{vdW} = \hbar\sum_{i<j}^{N}\frac{C_6}{|r_i-r_j|^6}\sigma_{rr}^i\sigma_{rr}^j$. Here, the transition operator is $\sigma_{\alpha\beta}^{j} = \left\vert \alpha\right\rangle\left\langle \beta\right\vert_{j}$($\alpha \neq\beta$) and the projection operator is $\sigma_{\alpha\alpha}^{j} = \left\vert \alpha\right\rangle\left\langle \alpha\right\vert_{j}$, $\Delta_{p(s)}$ is the one(two)-photon detuning, $C_6$ is the vdWs coefficient and $|r_i-r_j|$ the interatomic distance. $N= \rho V$ defines the number of atoms with $\rho$ the atomic density and $V$ the volume. Note that $1/\rho=4\pi R^3/3$ presents the occupied space of single atom with $R$ the average interatomic spacing. For the $j$th atom, the interaction part can be replaced by $\mathcal{U}_{vdW}=\hbar\sum_{j}^{N}\sigma_{rr}^{j}\sum_{i\neq j}\frac{ C_6}{|r_i-r_j|^6}\sigma_{rr}^{i}$ under the mean-field treatment \cite{Tony11}, by which the many-body interacting system is replaced by a model of one atom $j$ affected by the accumulated level shifts from other nearby exciting atoms. Note that the atom-atom correlations are neglected in this approximation.

To this end, the time evolution for $\sigma_{\alpha\beta}^{j}$ can be governed by the following motional equations
\begin{widetext}
\begin{eqnarray}
\dot{\sigma}_{gg}^{j} &=& i\Omega_p\sigma_{gm}^j - i\Omega_p^{*}\sigma_{mg}^{j}+2\gamma_{gm}\sigma_{mm}^j \\
\dot{\sigma}_{rr}^{j} &=& i\Omega_s\sigma_{mr}^j - i\Omega_s^*\sigma_{rm}^j \\
\dot{\sigma}_{gm}^j &=& (i\Delta_p - \gamma_{gm})\sigma_{gm}^j + i\Omega_p^*(\sigma_{gg}^j-\sigma_{mm}^j)+i\Omega_s\sigma_{gr}^j \\
\dot{\sigma}_{gr}^j &=& i(\Delta_s-s)\sigma_{gr}^j + i\Omega_s^*\sigma_{gm}^j - i\Omega_p^*\sigma_{mr}^j\\
\dot{\sigma}_{mr}^j &=&  (i((\Delta_s-s)-\Delta_p)-\gamma_{gm})\sigma_{mr}^j + i\Omega_s^*(\sigma_{mm}^j-\sigma_{rr}^j)-i\Omega_p\sigma_{gr}^j
\end{eqnarray}
\end{widetext}
where $\gamma_{gm}$ is the dephasing rate of $|g\rangle \to |m\rangle$ transition and $\Gamma_{m(r)}$ the spontaneous decay rates of $|m(r)\rangle$.
In deriving Eqs.(1-5), we have used the relations of $\Gamma_m = 2\gamma_{gm}$ and $\gamma_{mr}=\gamma_{gm}$ by considering $\gamma_{\alpha\beta}=(\Gamma_{\alpha}+\Gamma_{\beta})/2$, $\alpha,\beta\in(g,m,r)$ and $\Gamma_m\gg\Gamma_r$ \cite{Sheng13,Hning13}. Besides, $s=\sum_{i\neq j}\frac{C_6}{|r_i-r_j|^6}\sigma_{rr}^{i}$ characterizes the interaction induced energy shifts to state $\left\vert r_j\right\rangle$ caused by other exciting atoms within the ensemble. Typically these atoms exist beyond the blockade radius. 

We further replace the sum in $s$ with a spatial integral standing for all interactions of exciting atoms. In fact, only one atom is excited within a blockade radius $R_b$ and the separation $r$ between two exciting atoms meets $r>R_b$, so it is reasonable to introduce a short-range cutoff to the spatial integral at $R_b$ \cite{DeSalvo16,Han16}
\begin{equation}
s \approx \int_{R_b}^{\infty}\frac{C_6}{r^6}\sigma_{rr}\rho 4\pi r^2dr=\frac{4\pi C_6}{3R_b^3}\rho\sigma_{rr}
\label{ssn}
\end{equation}
where $\rho\sigma_{rr}$ represents the atomic exciting density in the ensemble. First, the steady state solutions $\sigma_{rr}$, $\sigma_{gm}^R$, $\sigma_{gm}^I$ can be formally expressed by assuming $\dot{\sigma}_{\alpha\beta}^j=0$, as
\begin{widetext}
\begin{eqnarray}
\sigma_{rr} = \frac{\Omega_p^2(\Omega_p^2+\Omega_s(x)^2)}{(\Omega_p^2+\Omega_s(x)^2)^2+2\Delta_p(\Delta_s-s)\Omega_s(x)^2+(\Delta_s-s)^2(\gamma_{gm}^2+\Delta_p^2+2\Omega_p^2)} \label{rr}\\
\sigma_{gm}^I = \frac{\gamma_{gm}(\Delta_s-s)^2\Omega_p}{(\Omega_p^2+\Omega_s(x)^2)^2+2\Delta_p(\Delta_s-s)\Omega_s(x)^2+(\Delta_s-s)^2(\gamma_{gm}^2+\Delta_p^2+2\Omega_p^2)} \label{gmI} \\
\sigma_{gm}^R = \frac{(\Delta_s-s)(\Omega_s(x)^2-\Delta_p(\Delta_s-s))\Omega_p}{(\Omega_p^2+\Omega_s(x)^2)^2+2\Delta_p(\Delta_s-s)\Omega_s(x)^2+(\Delta_s-s)^2(\gamma_{gm}^2+\Delta_p^2+2\Omega_p^2)} \label{gmR}
\end{eqnarray}
\end{widetext}
with $s$ a relevant parameter with respect to $\sigma_{rr}$ as in Eq.(\ref{ssn}).
The solution of $\sigma_{rr}$ is nonlinear and complicated. Consequently it is hard to estimate $s$ exactly.

Notice that the formal solution $\sigma_{rr}$ is a Lorentzian-like function with respect to $\Delta_s$ by considering $s=0$ in Eq. (\ref{rr}) (only one exciting atom) and $\Delta_p\ll\gamma_{gm}$, giving rise to the half-linewidth of single-atom Rydberg probability: $\omega = (\Omega_p^2+\Omega_s^2)/\sqrt{\gamma_{gm}^2+\Delta_p^2+2\Omega_p^2}$. To quantitatively estimate $s$, we find that the single-atom blockade radius can be roughly given by $R_b = (C_6/\omega)^{1/6} $ \cite{Qian13}. With definitions of $R$ and $R_b$, we can finally arrive at a reduced form of the approximated interaction $s$, which is \cite{Petrosyan11}
\begin{equation}
s = \frac{\omega}{\xi}\sigma_{rr}\approx\frac{\Omega_p^2}{\xi\sqrt{\gamma_{gm}^2+\Delta_p^2+2\Omega_p^2}}
\label{ss}
\end{equation}
where we used approximated $\sigma_{rr}$ obtained by the formal solution under the assumption of $s=0$ for single-atom excitation and $\Delta_p\ll\gamma_{gm}$, $\Delta_s<\Omega_s$. The coefficient $\xi=(R/R_b)^3$ treated as an adjustable parameter controlled by the atomic density $\rho$, which stands for the strength of simultaneous excitation of nearby atoms to the Rydberg state. $\xi>1$ means the blockade is imperfect. Replacing Eq.(\ref{ss}) into Eqs.(7-9) finally gives rise to the analytical expressions for the steady solutions $\sigma_{rr}$, $\sigma_{gm}^I$ and $\sigma_{gm}^R$. We note that the parameters $\sigma_{gm}^R$ and $\sigma_{gm}^I$ can directly arise a position-dependent polarization to the probe field with the probe susceptibility given by
\begin{equation}
\chi_p(x) = \eta(\sigma_{gm}^R+i\sigma_{gm}^I)
\end{equation}
and $\eta =2\rho\mu_{gm}^2/\hbar\epsilon_0\Omega_p $ \cite{Carvalho11}. 
Here, the real part $\eta\sigma_{gm}^R$ of susceptibility stands for the response of dispersion of the medium and the imaginary part $ \eta\sigma_{gm}^I$ for the medium absorption response.

\section{position-dependent two-photon detuning}

For an atomic medium modulated by the strong SW field along $x$ axis, the transmission function for the probe field can be solved from the propagation equation, given by
\begin{equation}
T(x) = e^{-\alpha(x)D+i\beta(x)D}
\end{equation}
where $\alpha(x) = (2\pi\eta/\lambda_p)\sigma_{gm}^I$ and $\beta(x) = (2\pi\eta/\lambda_p)\sigma_{gm}^R$ represent the amplitude and phase modulations. The atom-field interaction length $D=\zeta z_0$ with optical depth $\zeta$ in unit of $z_0 = \frac{\lambda_p}{2\pi\xi\eta}$ ($\lambda_p$ is the probe wavelength), characterizes the length of atom-field interaction along $z$ axis. By carrying out the Fourier transformation of $T(x)$ we can obtain the $n$th-order diffraction intensity of the probe field, given by
\begin{equation}
I_p(\theta_n) = |E_p(\theta_n)|^2\times\frac{\sin^2(M\pi\Lambda_{sx}\sin(\theta_n)/\lambda_p)}{M^2\sin^2(\pi\Lambda_{sx}\sin(\theta_n)/\lambda_p)}
\end{equation}
where the intensity in a single period is
\begin{equation}
E_p(\theta_n) = \int_{-\Lambda_{sx}/2}^{+\Lambda_{sx}/2}T(x)e^{-i2\pi nx }dx, 
\end{equation}
and $n = \Lambda_{sx}\sin\theta_n/\lambda_p$ is the diffraction order and $M$ is the number of grating period defined by the ratio between the beam width of $\Omega_p$ and the grating periodic number $\Lambda_{sx}$. The overall output $I_{out} = \sum_n I_p^{pk}(\theta_n)$ (n=0,$\pm 1$, $\pm 2$ ...) is defined by all maximal $n$th-order diffraction intensities $I_p^{pk}(\theta_n)$.

It is well known that, a normal Rydberg-EIG with symmetric diffraction intensities can be created, stemming from a spatial modulation by the strong coupling field $\Omega_{s}(x)$ to modify the dispersion $\eta\sigma_{gm}^R$ and absorption $\eta\sigma_{gm}^I$ of the medium \cite{Asghar16}. For $\eta$ is constant we will omit it and treat $\sigma_{gm}^R$($\sigma_{gm}^I$) as dispersion(absorption) in the following discussions. As a result, $\sigma_{gm}^R$ and $\sigma_{gm}^I$ show spatially-symmetric even functions, yielding symmetric diffraction patterns, {\it e.g.} see Fig.\ref{diffractionp}(a3). Observing exotic unidirectional diffraction (UD) requires the breakup of this symmetry, for that purpose, we introduce a suitable position-dependent adjustment to the two-photon detuning $\Delta_s(x)$, which can change the parity of the dispersion $\sigma_{gm}^R$ [Eq.(9)]. Similar spatial modulation to the light shift was considered in a lattice system manipulated by tuning the orientation of laser beams \cite{Wu14}. 

Here, for realizing an easy experimental control, we give an in-phase spatial modulation to $\Delta_s(x)$ as similar as $\Omega_s(x)$, with $\delta$ the modulation amplitude and $\Delta_{s0}$ the constant detuning,
\begin{equation}
\Delta_s(x) = \Delta_{s0}+\delta\sin(\pi x/\Lambda_{sx}) 
\end{equation}
which can be realized by the AC Stark effect to induce a periodic change of energy shift of state $|r\rangle$. In experiment, one can use extra spatially-modulated strong lasers for that purpose \cite{Alber94}. The position-dependent two-photon detuning $\Delta_s(x)$ accompanying with $\Omega_s(x)$ will lead to an anomalous change for the dispersion function $\sigma_{gm}^R(x)$, making it non-even. It is expected that, in the case of $\delta=0$ and $\Delta_s=\Delta_{s0}$, one creates a normal Rydberg-EIG because $\sigma_{gm}^R$ is exactly an even function. The resulting probe diffraction is expected to be diffracted uniformly into both positive and negative angles. However, in the case of $\delta\neq0$, $\Delta_s(x)$ will cause the key diffraction player: dispersion $\sigma_{gm}^R$ to be out of phase and an exotic UD may be observed.

\section{Numerical results and discussions}

\subsection{Unidirectional diffractions}

\begin{figure}
\centering
\includegraphics[width=3.5in,height=2.8in]{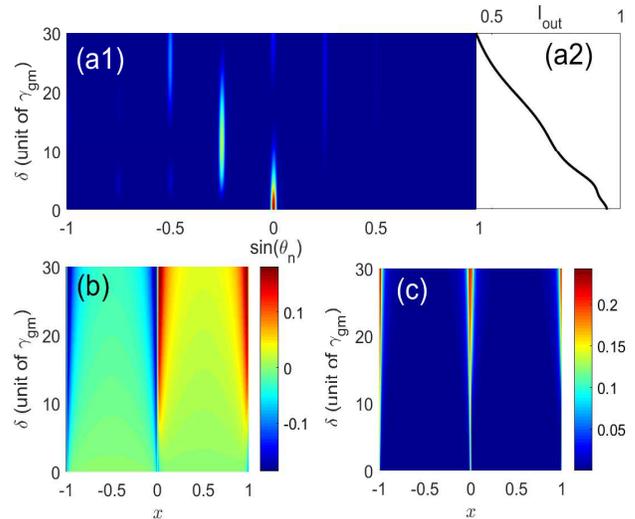}
\caption{(color online). (a1) The diffraction intensity $I_p(\theta_n)$ versus the modulation amplitude $\delta$ and the diffraction angle $\sin\theta_n$ for $\Delta_{s0}=0$. As increasing $\delta$, the diffraction intensity disperses into higher-order directions. (a2) The overall output $I_{out}$ versus $\delta$. (b) The values of dispersion $\sigma_{gm}^R$ and (c) of absorption $\sigma_{gm}^I$ versus $\delta$ and $x$. Other specific parameters are $\Omega_p=0.5\gamma_{gm}$, $\Omega_{s0}=22.5\gamma_{gm}$, $\Delta_p=0$, $M=10$, $\zeta=200$, $\xi=3.0$. }
\label{diffractionpn}
\end{figure}

To see the effect by the modulation from the detuning, we first plot the probe diffraction intensity $I_p(\theta_n)$ versus the modulation amplitude $\delta$ and the angle $\sin(\theta_n)$  in Fig. \ref{diffractionpn}(a1). The overall output intensity $I_{out}$ versus $\delta$ is shown in Fig. \ref{diffractionpn}(a2). In general, if $\delta\neq 0$ the diffraction is basically asymmetric and disperses into higher-order diffractions with the increase of $\delta$, accompanied by a slow decrease for the overall output $I_{out}$ due to the absorption effect. A special case is when $\delta=0$, the diffraction intensity totally gathers into the zeroth-order direction where only the SW coupling field plays roles, and it further disperses for $\delta\neq 0$ owing to the growing of dispersion affected by the periodic modulation from the two-photon detuning, giving rise to an UD grating. The underlying physics comes from the breakup of parity of the dispersion function, see Eq.(9) that is $\sigma_{gm}^R\propto(\Delta_s(x)-s)\Omega_s(x)^2$. When $\Delta_s(x)=\Delta_{s0}$, $\sigma_{gm}^R$ is even; otherwise it is modulated to be non-even with respect to $x=0$. 

Fig. \ref{diffractionpn}(b-c) show the variations of dispersion and absorption functions versus $\delta$ and $x$. Clearly, $\sigma_{gm}^R$ is modulated to contain positive and negative values as $\delta$ increases and meanwhile the increasing of $\sigma_{gm}^I$ persists, arising a continuous reduction to $I_{out}$. In a word, for a larger $\delta$, both dispersion $\sigma_{gm}^R$ and absorption $\sigma_{gm}^I$ are improved however only the parity of dispersion is significantly changed by the modulation.

\begin{figure}
\centering
\includegraphics[width=3.5in,height=3.0in]{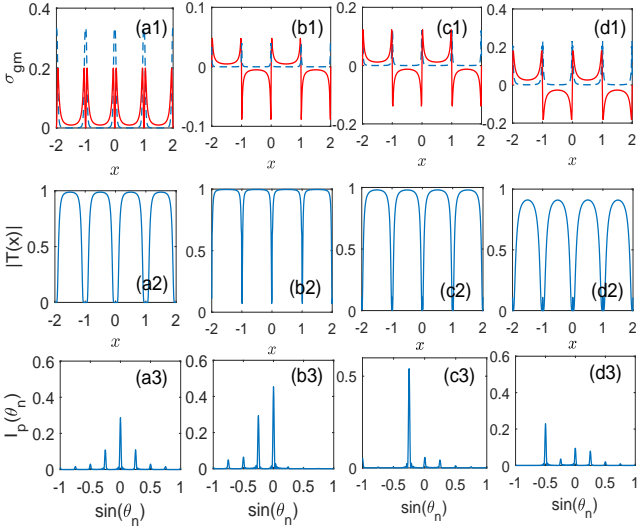}
\caption{(color online). Plots of dispersion $\sigma_{gm}^R$ (red solid), absorption $\sigma_{gm}^I$ (blue dashed), transmission $|T(x)|$, and diffraction pattern $I_p(\theta_n)$ are demonstrated in (a1-a3) for $\Delta_{s0}/\gamma_{gm}=10$ and $\delta=0$, in (b1-b3) for $\Delta_{s0}=0$ and $\delta/\gamma_{gm}=5$, in (c1-c3) for $\Delta_{s0}=0$ and $\delta/\gamma_{gm} = 12.5$, in (d1-d3) for $\Delta_{s0}=0$ and $\delta/\gamma_{gm} = 27$.}
\label{diffractionp}
\end{figure}

Specific results for $\delta/\gamma_{gm} = (0,5.0,12.5,27)$ are presented in Fig. \ref{diffractionp} where the patterns of  dispersion, absorption, transmission and diffraction intensities are respectively shown. As expected, for $\delta=0, \Delta_{s0}/\gamma_{gm}=10$[(a3)], the diffraction intensity $I_p(\theta_n)$ represents a perfect symmetric distribution with positive and negative diffraction angles owing to the even functions of $\sigma_{gm}^R$ and $\sigma_{gm}^I$, which are solely modulated by the SW coupling field. As $\delta$ is increased to 5.0$\gamma_{gm}$, the diffraction intensity turns to be anomalously unidirectional and only distributes in the range of negative angles. It is remarkable that this diffraction direction can be manipulated by changing the sign of modulation amplitude $\delta$ easily. A further increase of $\delta$ leads to the primary diffraction order transfers to negative first-order[(c3)] and second-order[(d3)] directions because of the growing of dispersion $\sigma_{gm}^R$. Meanwhile, the increasing of absorption induces a slight reduction to the transmission as well as the overall output.

\begin{figure}
\centering
\includegraphics[width=3.5in,height=2.6in]{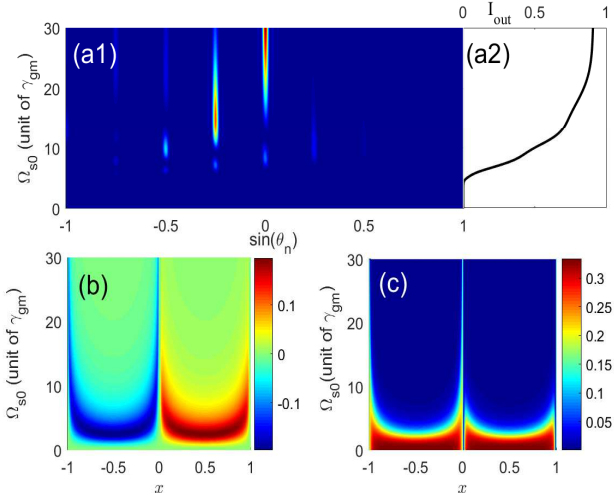}
\caption{(color online). As similar as Fig.\ref{diffractionpn} except that all physical quantities are shown with respect to $\Omega_{s0}$. Other specific parameters are $\delta=5.0\gamma_{gm}$, $\Omega_p=0.5\gamma_{gm}$.}
\label{diffractionstok}
\end{figure}

To search for the optimal conditions of a perfect UD [{\it e.g.}, Fig. \ref{diffractionp}(b3)], focusing on the competition between $\Delta_{s}(x)$ and $\Omega_s(x)$, we study the diffraction intensity $I_p(\theta_n)$ versus the variation of $\Omega_{s0}$ and $\sin(\theta_n)$ while keeping $\delta$ and $\Omega_p$ constant. It can be clearly seen that $I_p(\theta_n)$ [Fig. \ref{diffractionstok}(a1)] presents an opposite behavior with respect to that in Fig. \ref{diffractionpn}(a1), i.e. the diffraction intensity gathers into the zeroth-order direction as increasing $\Omega_{s0}$. The resulting overall output $I_{out}$ keeps growing which saturates towards $I_{out}\approx0.9$ as $\Omega_{s0}$ is sufficiently large $\Omega_{s0}\gg\Omega_p,\delta$. Actually, the essence for that can also be understood by the properties of dispersion and absorption. From Fig. \ref{diffractionstok}(b) it is observed that the dispersion $\sigma_{gm}^R$ is critically odd with a big amplitude as $\Omega_{s0}\to 0$ but it turns to be non-odd with decreasing amplitudes for a larger $\Omega_{s0}$. That is to say, a big modulation by the coupling field will lead to a convergence of the diffraction but the UD pattern persists due to $\delta\neq 0$. Accordingly, the absorption $\sigma_{gm}^I$ of the medium persists an even function however with rapidly-decreasing amplitude and width as $\Omega_{s0}$ increases, perfectly agreeing with the behavior of $I_{out}$, since a big absorption of medium represents that the diffraction is weak, and vice versa.

\begin{figure}
\centering
\includegraphics[width=3.5in,height=3.2in]{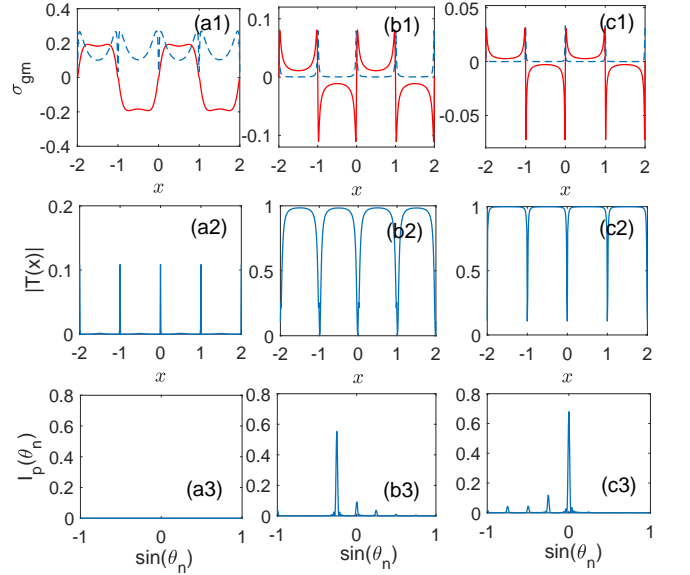}
\caption{(color online). Parameters $\sigma_{gm}^R$, $\sigma_{gm}^I$, $|T(x)|$, $I_p(\theta_n)$ are respectively shown for $\Omega_{s0}/\gamma_{gm}=3.0$ in (a1-a3); for $\Omega_{s0}/\gamma_{gm}=15$ in (b1-b3); for $\Omega_{s0}/\gamma_{gm}=30$ in (c1-c3). }
\label{diffractionstokn}
\end{figure}

Figure \ref{diffractionstokn} shows specific results of dispersion $\sigma_{gm}^R$, absorption $\sigma_{gm}^I$, transmission $|T(x)|$ and diffraction pattern $I_p(\theta_n)$. It is observed that, at $\Omega_{s0}=3.0\gamma_{gm}$, the dispersion and absorption are modulated to be broadened and large which give rise to poor transmission and diffraction intensity. As increasing $\Omega_{s0}$ to 15$\gamma_{gm}$, the improvement of transmission significantly enhances the intensity of diffraction, leading to a dominant negative first-order diffraction with its efficiency as high as $\sim 0.6$. It is expected that, for $\Omega_{s0}=30\gamma_{gm}$ the transmission is even larger due to the suppression of absorption, which yields a perfect UD with its maximal intensity located at the zeroth-order direction.

By comparing the results from two modulations, we summarize that to realize a perfect UD like cases of Fig.\ref{diffractionp}(b3) and Fig. \ref{diffractionstokn}(c3), it requires the condition of $\Omega_{s0}\gg\delta\gg \Omega_p$. Here, the EIG effect with $\Omega_{s0}\gg \Omega_p$ allows a periodic phase and amplitude modulation to the dispersion and absorption of the medium respectively. And the newly-introduced in-phase modulation for the two-photon detuning serves as a non-trivial control knob that can break the parity of dispersion, arising novel unidirectional diffractions. For a perfect UD, its modulation amplitude $\delta$ should be moderate.

\subsection{Controllable higher-order diffractions}

In a Rydberg-EIG, the influence of the vdWs shift $s$ directly related to the population of $|r\rangle$, will have a special contribution to the diffraction. Intuitively speaking, $s$ only introduces an energy-level shift due to interactions to the Rydberg state, which is quite different from the role of the position-dependent modulation $\Delta_s(x)$. From its definition [Eq.(\ref{ss})], it can be seen that an easy way to vary $s$ is experimentally controlling the average interatomic distance $R$, allowing the ratio $\xi=(R/R_b)^3$ varying in a large range. However, we note that the length unit $z_0$ also depends on $\xi$, so varying $\xi$ causes a consistent change for the interaction length $D$. To this end, we will study this mutual effect implemented by the vdWs interaction $s$ and the interaction length $D$.

\begin{figure}
\centering
\includegraphics[width=3.5in,height=2.5in]{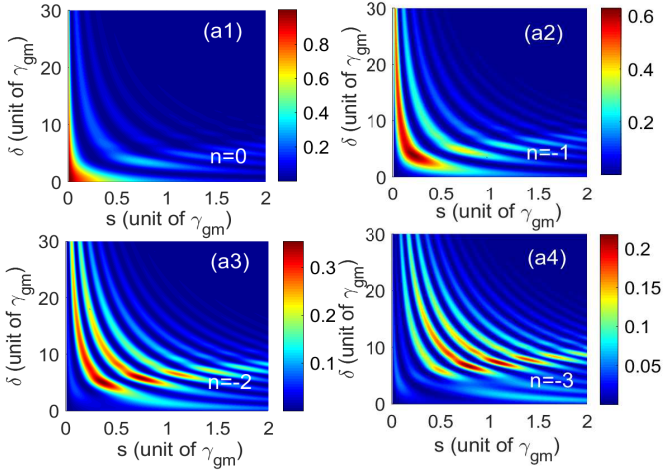}
\caption{(color online). The maximal diffraction intensity $I_p^{pk}(\theta_n)$ of $n$th-orders versus the vdWs interaction $s$ and the modulation amplitude $\delta$ for $n=0,-1,-2,-3$, respectively. }
\label{mutualeff}
\end{figure}

\begin{figure}
\centering
\includegraphics[width=3.5in,height=4.2in]{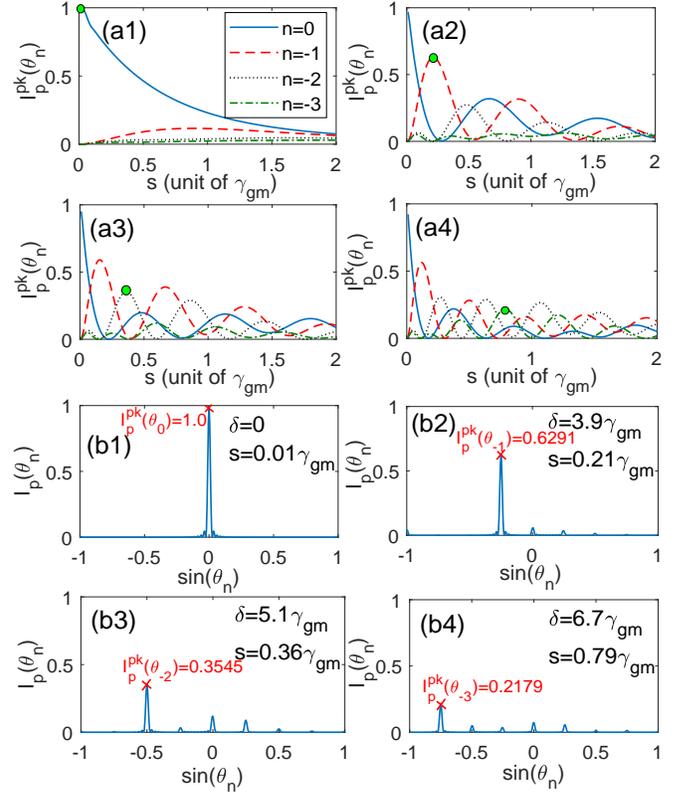}
\caption{(color online). The peak diffraction intensity $I_p^{pk}(\theta_n)$ versus the vdWs interaction $s$ for $n=0$ (blue solid), $n=-1$(red dashed), $n=-2$(black dotted) and $n=-3$(green dash-dotted) by using (a1) $\delta=0$, (a2) $\delta=3.9\gamma_{gm}$, (a3) $\delta=5.1\gamma_{gm}$ and (a4) $\delta=6.7\gamma_{gm}$. (b1-b4) The corresponding diffraction patterns $I_p(\theta_n)$ versus $\sin(\theta_n)$ for $n=0,-1,-2,-3$ where the values of $I_p^{pk}(\theta_n)$, $\delta$ and $s$ are denoted. Here, $\Omega_p=0.5\gamma_{gm}$, $\Omega_{s0}=22.5\gamma_{gm}$.}
\label{mutualeffn2}
\end{figure}

Fig. \ref{mutualeff} represents the peak intensity $I_p^{pk}(\theta_n)$ of $n$th-order diffractions versus $s$ and $\delta$ for $n=0,-1,-2,-3$ respectively. In general, $I_p^{pk}(\theta_n)$ is very sensitive to the values of $s$ and $\delta$, presenting a significant oscillating behavior. Specifically, in the absence of modulation $\delta\to 0$, $I_p^{pk}(\theta_n)$ is expected to continuously decrease with $s$, satisfying $I_p^{pk}(\theta_0)>I_p^{pk}(\theta_{-1})>I_p^{pk}(\theta_{-2})>I_p^{pk}(\theta_{-3})$, as shown by Fig.\ref{mutualeffn2}(a1). Similar results have been verified in Ref.\cite{Asghar16} due to the fact that the EIT effect does not work when state $|r\rangle$ is largely shifted, giving rise to a poor higher-order diffraction. 


More interestingly, for a non-zero $\delta$, $I_p^{pk}(\theta_n)$ exhibits a rapidly-oscillating behavior with the vdWs interaction $s$, which is exactly different from the previous findings that the diffraction intensity continuously decreases with $s$. This significant oscillation with $s$ comes from the mutual effect between $s$ and $D$ because $D$ also increases with $s$ i.e. $\xi^{-1}$. Increasing $D$ leads to an enhancement for the dispersion and absorption modulation depth, which further transfers the probe diffraction energy into higher-order directions. For the same reason, by increasing $\delta$, the peak diffraction intensity is also transferred into higher-order directions owing to the enhanced-dispersion with $\delta$, as indicated by Fig.\ref{diffractionpn}(b). 
Therefore, a grating with the enhanced higher-order diffraction can be obtained when the values of $\delta$ and $s$ are both optimally selected.

Fig. \ref{mutualeffn2}(a1-a4) demonstrate the variations of the $n$th-order peak diffraction intensity versus $s$ for  $\delta/\gamma_{gm}=(0,3.9,5.1,6.7)$. It is observed that all $n$th-order intensities decrease with $s$ for $\delta=0$. However, 
once $\delta$ is non-zero, the probe diffraction energy persists transferring into the higher-order directions along the orientation of $I_p^{pk}(\theta_0)\to I_p^{pk}(\theta_{-1})\to I_p^{pk}(\theta_{-2})\to I_p^{pk}(\theta_{-3})$ with the enhancement of $s$, presenting a $s$-dependent oscillation. Similar oscillations between first- and second-order diffraction intensity with respect to the interaction length was found by Ref.\cite{Bozorgzadeh18}. 

In Fig. \ref{mutualeffn2}(b1-b4) we show the distributions of $n$th-order diffraction intensity by suitably adjusting $\delta$ and $s$, allowing the roles of the zeroth-order, the negative first-order, second-order and third-order diffractions are dominant, as denoted with green circles in (a1-a4). By controlling $\delta$ and $s$, we are able to obtain enhanced higher-order diffractions and even the negative third-order diffraction can achieve as high as $I_p^{pk}(\theta_{-3})\approx0.2179$ when $\delta$ and $s$ are enough large to enhance the dispersion.


 \section{Experimental realization}

Experimental consideration for the implementation of a novel Rydberg-EIG scheme is performed in an ultracold atomic ensemble of $^{87}$Rb atoms with energy levels $\left\vert g\right\rangle = 5S_{1/2}\left\vert F=2,m_F=2\right\rangle$, $\left\vert m\right\rangle = 5P_{1/2}\left\vert F=3,m_F=3\right\rangle$, $\left\vert r\right\rangle = 62S_{1/2}$. The spontaneous decay of $\left\vert m\right\rangle$ is $\Gamma_e/2\pi = 6.1$MHz, giving to the dephasing rate $\gamma_{gm}/2\pi=3.05$MHz. The dimensionless value $\eta$ is $7.18\times10^{-4}$ by using $\rho=5\times10^{10}$cm$^{-3}$, $\mu_{gm}=2.534\times10^{-29}$Cm, $\xi=3.0$, yielding the length unit of the system $z_0=1.848\mu$m, the optical depth $\zeta=200$. The resulting interaction length is expected to be $D=369.6\mu$m (equivalent to the values used in {\it e.g} ref.\cite{Liuol16}), which can be varied in a large range by the average distance $R$. The probe wavelength is $\lambda_p = 0.25\mu$m, by which we can simply assume the spatial period of the grating is $\Lambda_{sx}=4\lambda_p=1.0\mu$m.
In the simulations we employ a weak probe laser $\Omega_p/2\pi=1.525$MHz and a wide-range adjustment for the strong coupling field $\Omega_{s0}/2\pi\in(0,91.5]$ MHz in a reasonable range. The auxiliary spatial modulation by Stark shifts from an off-resonant laser field induces a comparable modulation $\Delta_s(x)$ with the amplitude $\delta/2\pi\in(0,91.5]$ MHz \cite{Hang13}. The typical timescale to reach the steady state solution is about $\sim \mu s$ as same as used in most experiments. Finally, with an optimum control for parameters we can realize unidirectional higher-order diffractions, see some optimal results summarized in Table \rm{I}. It is clear to see that in a strong-blockade environment ($R/R_b<1$), it is easier to obtain a controlled higher-order diffraction grating with the aid of competitive modulations of $\delta$ and $\Omega_{s0}$.

\begin{table}[tbp]
\begin{tabular}{c|c|c|c|c|c}
\hline\hline
order & \multicolumn{4}{c}{key parameters} & \multicolumn{1}{c}{peak intensity} \\
\hline 
 n & $\delta$(MHz) & $R/R_b$ & $s$(MHz) & $D$(mm) & $I_p^{pk}(\theta_n)$ \\ 
 \hline
0 & 0 & 2.73 & 0.192 & 0.054 & 1.0  \\ 
-1 & 74.70 & 0.99 & 4.022 & 1.140 & 0.6291 \\ 
-2 & 97.69 & 0.83 & 6.895 & 1.956 & 0.3545 \\
-3 & 128.33 & 0.64 & 15.132 & 4.289 & 0.2179  \\
\hline\hline
\end{tabular}%
\caption{According to Fig.\ref{mutualeffn2}(b1-b4), the optimal values of peak $n$th-order diffraction intensity $I_p^{pk}(\theta_n)$ are summarized with relevant parameters required $\delta$, $R/R_b$, $s$, $D$. The laser fields are $\Omega_p/2\pi=1.525$MHz, $\Omega_{s0}/2\pi=68.63$MHz.}%
\label{summ}
\end{table}

\section{Summary}

We investigate a scheme for realizing an exotic unidirectional Rydberg-EIG in a three-level cascade system by implementing a position-dependent two-photon detuning to break the parity of the dispersion of medium. In a normal EIG, the strong SW coupling field brings a spatially-periodic modulation to the refractive index of the medium, which uniformly diffracts the weak TW probe field into positive and negative directions. Here, owing to the parity breaking of dispersion, the transmission function is also affected, leading to an unidirectional diffraction pattern (only negative-angle direction is diffracted that depends on the sign of modulation amplitude $\delta$). We find that it is feasible to design an atomic grating with perfect unidirectional diffractions by using an appropriate modulation amplitude for the two-photon detuning, {\it i.e.} $\Omega_{s0}\gg \delta\gg \Omega_p$. Furthermore, differing from the previous results that increasing vdWs interaction causes a continuous damping to the $n$th-order diffraction intensity, here, with the increase of vdWs interaction, it can be shown that the maximal $n$th-order diffraction intensities represent anomalous oscillations and are transferred into higher-order diffractions due to the mutual interplay between the vdWs interaction and the interaction length of the medium, which may provide more prospectives to realize enhanced higher-order diffractions even in the case of strong blockade.

Our study offers a new approach to improve the intensity of unidirectional higher-order diffraction, paving the avenue for designing new Rydberg quantum devices such as all-optical quantum switches, large-angle all-optical splitter, {\it etc.} based on the technique of Rydberg-EIG.

\bigskip

\acknowledgements

This work is supported by the NSFC under Grants No. 11474094, No. 11604086 and No. 11104076,
by the Science and Technology Commission of Shanghai Municipality under Grant No. 18ZR1412800, 
the Specialized Research Fund for the Doctoral Program of Higher Education No. 20110076120004.

\appendix

\end{document}